\documentclass[aps,pre,superscriptaddress,reprint]{revtex4-1}
\usepackage{amsmath,amssymb}
\usepackage{graphicx}
\usepackage{dcolumn}
\usepackage{bm}
\usepackage{nameref}
\usepackage{hyperref}
\usepackage{courier}
\usepackage{braket}
\usepackage[rightcaption]{sidecap}
\usepackage{float}
\usepackage[usenames,dvipsnames]{xcolor}

\usepackage[most]{tcolorbox}
\newtcolorbox{highlighted}{colback=yellow,coltext=red,breakable}
\usepackage{lipsum}
\usepackage[normalem]{ulem}

\hypersetup{
 colorlinks   = true,	
  urlcolor     = blue, 
  linkcolor    = blue, 
  citecolor   = blue 
}
\begin{document}
\newcommand{\ve}{\varepsilon}
\title{Beyond Dry Flocking: Shape-Dependent Hydrodynamic Alignment
}
\author{Abhishek Thakur}

\affiliation{Department of Physics, Indian Institute of Technology Bombay, Powai, Mumbai 400 076, India}

\author{Taniya Khatun }
\affiliation{Department of Physics, Indian Institute of Technology Bombay, Powai, Mumbai 400 076, India}
\affiliation{Present affiliation: Department of Mathematics, Indian Institute of Technology Ropar, Rupnagar, Punjab 140001, India}

\author{Soumyadeep Kundu}
\affiliation{Department of Physics, Indian Institute of Technology Bombay, Powai, Mumbai 400 076, India}

\author{Nitin Kumar}
\affiliation{Department of Physics, Indian Institute of Technology Bombay, Powai, Mumbai 400 076, India}

\author{P. Parmananda}
\affiliation{Department of Physics, Indian Institute of Technology Bombay, Powai, Mumbai 400 076, India}

\author{Anirban Sain}
\email[
asain@phy.iitb.ac.in]{}
\affiliation{Department of Physics, Indian Institute of Technology Bombay, Powai, Mumbai 400 076, India}

\begin{abstract}
\noindent
Flocking results from dynamical alignment among
neighbors in a collection of active self propelling objects. The intervening medium, typically a fluid, however, is not considered to be important for alignment, except acting as a momentum sink. The flocking system is therefore described as momentum nonconserving ``dry" active matter. Here we investigate a confined system of self-propelled Marangoni swimmers where feedback from the fluid medium, which crucially depends on the shape of the swimmers, influences their alignment. Using up to thirty, partially camphor coated, triangular paper boats-the polar swimmers,  we uncover a disorder to order, and an unusual, re-entrant, order to disorder transition, as functions of  camphor concentration and swimmer density, respectively.
We reproduce these transitions by formulating a reaction–diffusion model for the camphor field, coupled to the swimmer dynamics, where propulsion forces arise from surface tension gradient. 
The shape dependent hydrodynamic interaction, discussed here, could be central to efficient flocking in living organisms too. 
\end{abstract}.

\maketitle

Collective unidirectional motion or flocking, among birds, fish and animals, for long have been believed to be driven by strategy that enhances survival and fitness. But over the last thirty years laboratory experiments with collection of ``active" entities \cite{marchetti2013hydrodynamics}, that convert energy into directed motion, have successfully reproduced rich variety of collective behaviors, including flocking \cite{vicsek1995novel,bricard2013emergence,marchetti2013hydrodynamics,kumar2014flocking,elgeti2015physics,weber2013long, sar2023flocking}, milling \cite{couzin2002collective,huang2024collective}, clustering \cite{zottl2016emergent,bechinger2016active}, and phase separation\cite{zottl2016emergent,bechinger2016active,su2023motility,o2023introduction,kumar2019trapping}.
Examples of such active systems are self-propelling colloids \cite{bricard2013emergence}, vibrated polar disks \cite{weber2013long}, rods \cite{kumar2014flocking, kumar2019trapping}, robots \cite{molodtsova2025micellization, paramanick2025spontaneous,thakur2026resetting}, living system \cite{chao2026selective} and Marangoni swimmers \cite{suematsu2010collective, gilbert2006huddling,  heisler2012swarming, maitra2025marangoni}. \\

In this letter we ask to what extent does shape of the active entities matter for flocking. After all, stability of alignment between a pair of swimmers depends on hydrodynamic interaction which in turn depends on their shape. 
Other than persistent self-propulsion, flocking also requires a tendency for the entities to reorient upon collision (encounter). However, unless most of these encounters are ``favorable", leading to alignment between neighbors, a flock cannot emerge. Non-favorable encounters act as mis-aligning, random forces or noise. 
For example, in Viscek type models \cite{vicsek1995novel} for point particles, flocking occurs at low noise and high density. Density and propulsion speed both are important as they increase frequency of encounters, and in turn the Peclet number, often a criterion for ordering. In addition, physical confinement can also increase frequency of encounters and can even compensate for low density, as we show for our system of Marangoni swimmers. The focus of our paper is how swimmer shape
influence probability of ``favorable" encounters and  whether it can render stability to an aligned pair exploiting 
feedback from the fluid medium. \\ 




Our search for flocking behaviour with Marangoni swimmers crucially depends upon achieving precise control over the self-propelling, persistent, straight line motion of a single swimmer. Early experiments on dry granular particles \cite{kumar2019trapping, kumar2014flocking} or Marangoni swimmers \cite{dey2025run} had established elongated, tapered geometry to be strong candidates for producing persistent motion and alignment. Different types of single particle motion, including linear\cite {nakata1997self,dey2025confinement}, circular\cite{loffler2023new,koyano2019rotational,nakata2016selection,maitra2025marangoni} and oscillatory\cite{sharma2020rotational,monish2024stationary,tiwari2020periodic,suematsu2015synchronized,koyano2021imperfect} had also been discovered using different particle geometries.
Subsequent work by Suematsu \textit{et al.} \cite{suematsu2010collective} and  Heisler \textit{et al.} \cite{heisler2012swarming} also
reported swarming behavior in restricted geometry.
However, recent work by Song \textit{et al.} \cite{song2023pen} introduced a novel approach using a pen to draw camphor patterns that can guide Marangoni swimmers with enhanced precision, offering a degree of controllability previously unattainable. Using this approach we had recently 
reproduced bacterial-like run-and-tumble motion for single Marangoni swimmers \cite{dey2025run}. The present work is aimed at flock formation with similar type of swimmers with highly tunable propulsion dynamics. \\

\begin{figure}
    \centering
    \includegraphics[width=1.0\linewidth]{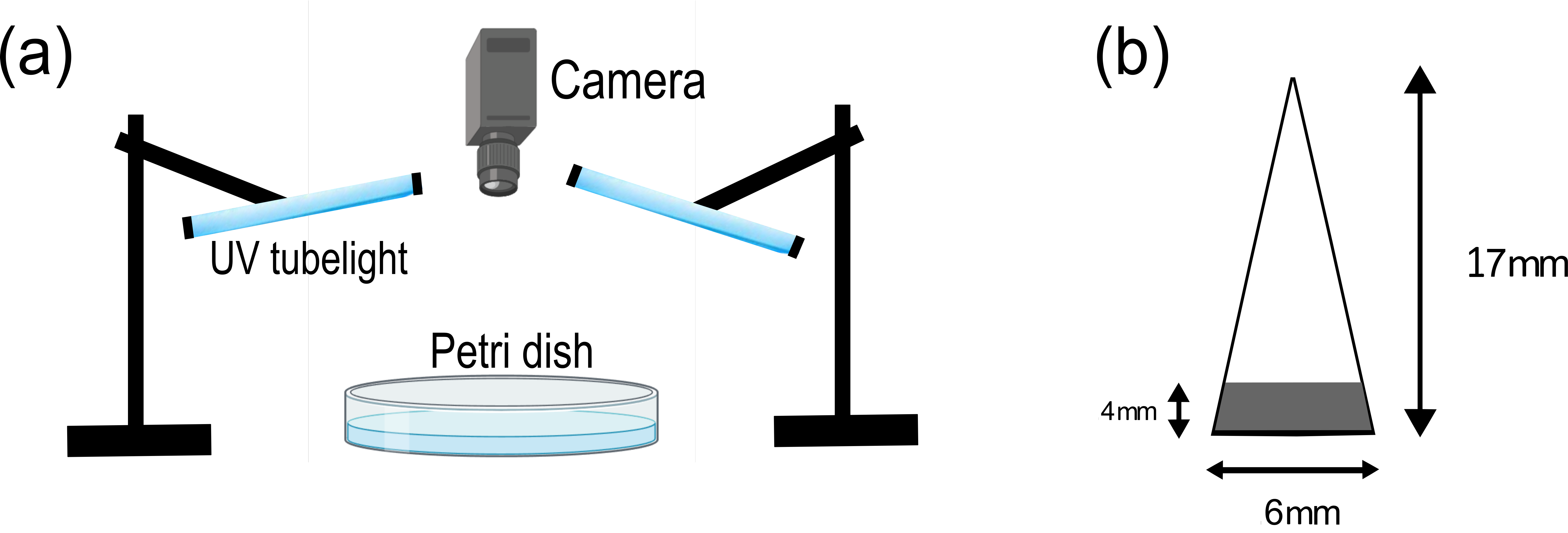}
    \caption{(a) Schematic view of the experimental setup. (b) Configuration of a single swimmer, with
    the gray region indicating the camphor-ethanol coated area.}
    \label{fig:exp_schme}
\end{figure}


In this work, we experimentally investigate (for experimental details see supplementary information (SI) section III) the collective dynamics of self-propelled swimmers (SPS) confined within a circular boundary (Fig.\ref{fig:exp_schme}). Our main result is discovery of two distinct transitions, one disorder to order and another order to disorder, by tuning the camphor–ethanol concentration ($M$), and number of swimmers ($N$), respectively.   The transition with respect to $M$ is shown in Fig.\ref{fig:fig2} where flocking corresponds to the ordered state. The transition with respect to $N$ is shown in section VI Fig.7 of (SI). 
\par

\begin{figure*}
    \centering
    \includegraphics[width=1.0\linewidth]{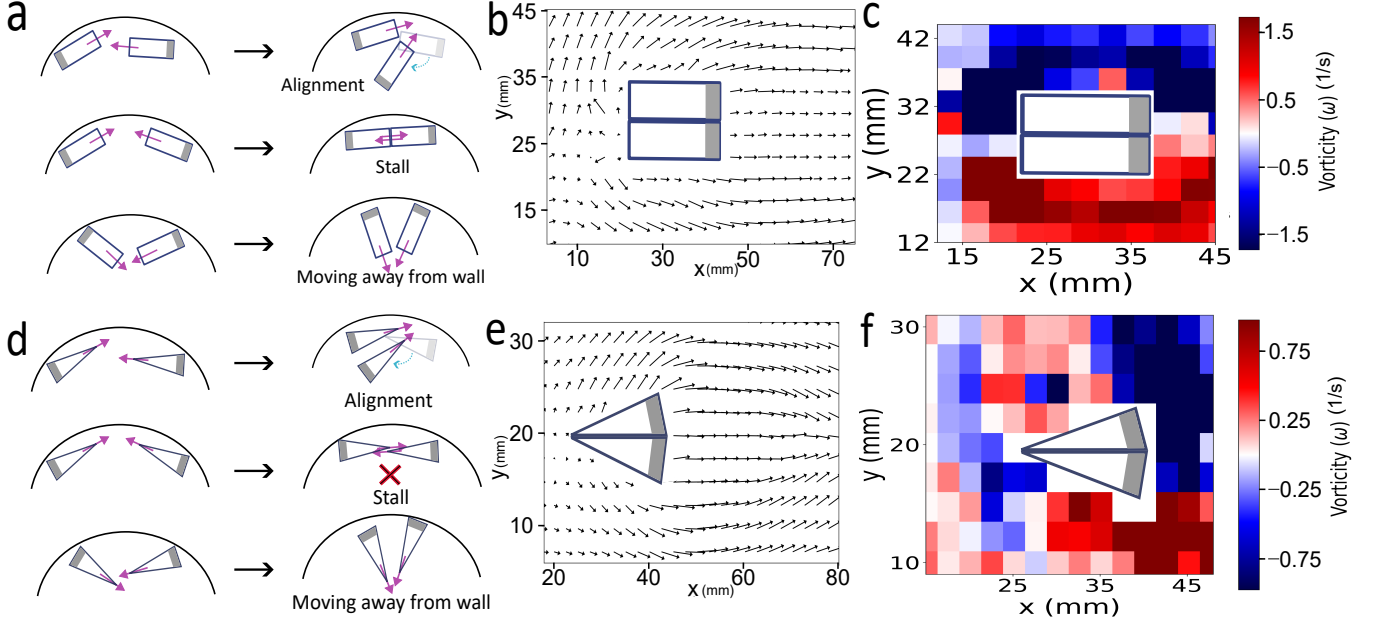}
    \caption{Role of swimmer geometry, rectangular 
    in (a) and triangular in (d), show influence 
    of shape on collision-induced alignment and stalling. Low probability of head on collisions in triangles avoids stalling. 
    The grey color at the back of the swimmers indicate the camphor releasing patch, driving propulsion.
    (b,e) Fluid velocity field, generated by an aligned pair of swimmers (in the swimmer's frame of reference), captured using PIV of  added tracer particles. (c,f) are the corresponding vorticity fields, derived from (b,e).  Relatively higher vorticity near the blunt moving front in (c) destabilizes the rectangle pair. During PIV, velocity could not be obtained at the pixels that had overlap with the object mask.
    }
    \label{fig:fig_1}
\end{figure*}

 Here we emphasize that confinement alone is insufficient to generate flocking; we performed control experiments using swimmers of different geometries (SI section II), for example, circular (SI section II Fig.a) and rectangular (SI section II Fig.b).
 Although these shapes exhibited boundary-following motion due to circular confinement, none were able to sustain long-term collective alignment or coherent flocking. This highlights the importance of shape and how it influences ``favorable" encounters. For circular confinement, it turns out that frequent occurrence of head-to-head collisions between swimmers along the boundary, leads to mutual stalling, see Fig.\ref{fig:fig_1}a for rectangular swimmers. Such stalling events deter flock formation by interrupting persistent motion of a globally aligned state. 
 In contrast, for triangular swimmers (SI section II Fig.c), the probability of direct head-to-head collisions along the boundary is extremely low due to their pointed head (see Fig.\ref{fig:fig_1}d). Instead of stalling, most encounters in fact promote reorientation and alignment (Movie1), helping to maintain persistent, coherent motion over long times.\\

While flocking starts with alignment between two self-propelling neighbors, persistence of this alignment over macroscopic time scales is crucial. 
We therefore check for hydrodynamic stability of an aligned pair of
swimmers, by studying the destabilizing torques produced by the fluid flow. For this purpose we construct a composite 
swimmer which mimics two aligned swimmers (Movie2), as shown in Fig.\ref{fig:fig_1}(b and e)
and study the resulting velocity ${\bf v}(x,y)$ and vorticity fields
$\omega (x,y)=\Big(\frac{\partial v_x}{\partial y}-\frac{\partial v_y}{\partial x}\Big)$ of the fluid,
for rectangle and triangle
pair. Fig.\ref{fig:fig_1} shows the resulting fluid velocities (b,e) and the
vorticities (c,f) of the rectangle and triangle pairs.  Note that, near the front of the swimmer (nose) the vorticity for the triangle pair shows much lower values compared to that for the rectangular pair (while their propulsion speeds are similar). Even near the tails the vorticity
magnitudes for the triangle pair are half that of the rectangle pair. This has two implications: 
1) As vortical fluid motion near a sharp corner requires centripetal force (radial pressure gradient) on the fluid, the reaction force on the solid body is directed radially outward from the body producing destabilizing torques, leading to separation of the pair. 2) Higher destabilizing shear stress on the sides of the rectangular pair. Clearly, the rectangular pair is unstable. In summary, the streamlined shape of the nose of the triangle pair leads to  stability. Note that the low pressure wake zone of the triangle and the rectangle-pair have comparable sizes, which in principle can 
draw in more swimmers and grow the flock. However,
if alignment of the leading pair is unstable then the flock is doomed. 
\begin{figure}
    \centering   \includegraphics[width=1.0\linewidth]
 {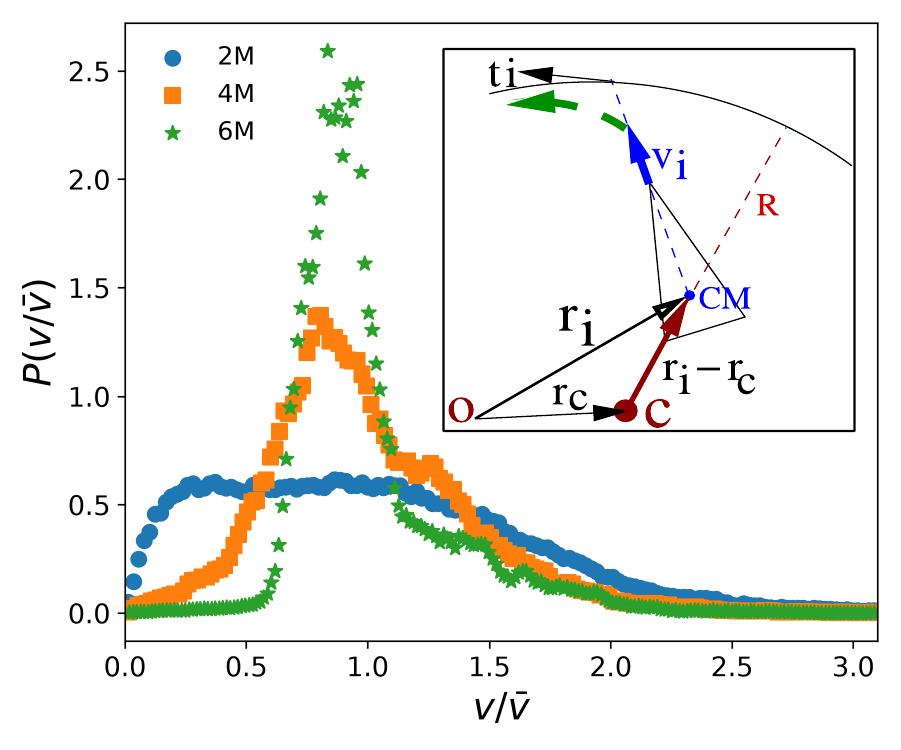}
    \caption{Probability distribution of normalized speeds $v_i/ \bar{v}$ for a system of $i =$ 1 to 30 swimmers, at three different camphor concentrations (see legends). The velocity distribution 
    become sharper at higher camphor concentration, indicating stronger ordering. Inset shows schematic of boundary-induced alignment in a circular confinement (radius $R$, center at $C$). 
A swimmer, with its center of mass (CM) at $\mathbf{r}_i$ and
moving with velocity ${\bf v}_i$, 
experiences a torque that aligns its orientation $\theta_i$ with the local 
tangential direction $\hat t_i$ at the boundary. The magnitude of the torque sharply decay away from the wall following a
short range gaussian of width $\sigma_{\mathrm{wall}}$. This wall aligning torque approximately models Cheerios effect.}
    \label{fig:align_torque}
\end{figure}

In order to quantify the degree of collective order, we introduce a global order parameter, 
\begin{equation}
\Phi(t) = \frac{1}{N}\left|  \sum_{i=1}^{N} \hat{\mathbf{r}}_i(t) \times \hat{\mathbf{v}}_i(t) \right|\;,
\label{eq.OP}
\end{equation}
defined as the average normalized angular momentum of the swimmers about the center of the Petri dish. Here, $\hat {\bf r}_i$ is the unit vector along ${\bf r}_i$, the position 
of the $i$-th swimmer (actually its center of mass) with respect to the center of the dish, marked as `C' in Fig.\ref{fig:align_torque}(inset). Similarly, $\hat {\bf v}_i$ is the corresponding unit vector along the velocity of the $i$-th swimmer. 
To analyze the dynamics the swimmers positions were tracked using the TrackMate plugin in FIJI (ImageJ) \cite{schindelin2012fiji,ershov2022trackmate,tinevez2017trackmate}, which allows for efficient detection of the center of mass  of each particle. The velocities were computed comparing  successive snapshots, $1/60$ seconds apart. \\

Using the time-averaged value of this order parameter \(\langle \Phi \rangle\), which measures the degree of collective circulation, 
we could detect two transitions: 1) disorder to order (Fig.\ref{fig:fig2}) with increase in camphor concentration, and 2) order to disorder (SI section VI) with increase in the 
swimmer number. Across these transitions we could identify three distinct dynamical states, (1) a solid flocking state ($\Phi\simeq 1$) (Movie3), (2) an intermediate flocking state ($1>\Phi>0.6 $) (Movie4), and (3) a disordered state ($\Phi<0.6$) (Movie5). 
In the solid flocking state the swimmers maintain fixed relative positions, forming stable pairs both side wise as well as along their front and rear, resembling a rigid body.  
In the intermediate flocking state, partial alignment is observed, with neighbor exchange 
events occurring intermittently. Also local clusters dynamically form and dissolve, maintaining a small life time. In the disordered
state, uncoordinated motion results in lack of
global order and slow dynamics. 


In Fig.\ref{fig:align_torque} we present another quantification of the collective dynamics, namely, the probability distribution of the normalized swimmer speed $v_i/\bar{v}$, at different camphor concentrations. At 2M,
the distribution is broad, indicating significant heterogeneity in speed, characteristic of a disordered state. At 4M, 
the distribution becomes narrower, reflecting partial alignment among swimmers. Finally, at 6M, the distribution exhibits a sharp peak around unity, signifying coherent, rigid-body-like motion.\\

Some of these features can be understood in terms of Péclet number (ratio of the diffusive and advective time scales) which depends on the camphor concentration ($C$) and the number of particles ($N$), as we explain below. The key points to note here are, (a) at higher $C$, propulsion velocity of the swimmers increases as the surface tension gradient which varies over the size of the swimmer (between its head and tail) increases. (b) Movement of the swimmers creates a velocity gradient which is responsible for fast dispersion  of the released camphor. (c) At higher $N$, the camphor gradient decreases as there are many release points (sources) and the area of influence of each swimmer decreases as these areas start to overlap. \\

To get an estimate of the typical velocities here, we note that the Marangoni stress at the interface is the main driver of the fluid flow. The flow in turn carries (advects) the swimmers. We therefore
compare the viscous stress 
${\sigma} = \eta(\nabla \mathbf{u} + \nabla \mathbf{u^T})$, where ${\bf u}$ is the velocity 
of the swimmer, with the Marangoni stress at the
interface
$\boldsymbol{\sigma_s}.\hat n$, where $\hat n$ is the surface normal. Marangoni stress is associated with gradients in surface tension $\gamma$, and decreases with camphor concentration $C$, approximately as, $\gamma ({\bf r}) =\gamma_0 -\alpha C({\bf r})$. Thus,  
$\boldsymbol{\sigma_s}.\hat n = -\alpha \nabla  C$, $\alpha$ being a  proportionality constant.\\

Equating the two stresses, leads to a characteristic interfacial velocity scale $U$ :
\[
\frac{\eta U}{\delta_z} \sim \frac{\alpha  \Delta C}
{\ell_p}\;,\; \mbox{and}\;\;\;U \sim \frac{\alpha \Delta C \, \delta_z}{\eta \ell_p}\;,
\]
where $\Delta C$ is the difference of camphor concentration across the length of the swimmer 
$\ell_p$. Further, $\delta_z$ is the interface thickness with viscous shear stress ($\eta \partial U/\partial z$) that balances the Marangoni stress. We assume that the interface thickness $\delta_z$ acquires a steady state value due to continuous injection and  evaporation of camphor into it and out of it,
respectively. Further,
$\ell_N$ is the characteristic separation between the sources (swimmers), given by $\sqrt{L^2/N}$, with $L$ the system size and $N$ the number of swimmers. Here $l_N$ defines the domain of influence of each boat. As $l_N$ decreases (at larger $N$) the camphor density gradients around the swimmers start to overlap and hence the resulting surface tension gradient becomes weak. \\

Estimating Péclet number as the ratio of the diffusive and advective time scales, we get
\begin{equation}
Pe \equiv \frac{\tau_{\text{diff}}}{\tau_{\text{adv}}} \sim \frac{\ell_N^2}{D} \cdot \frac{U}{\ell_N}\sim
\frac{\alpha \Delta C \delta_z \ell_N}{\eta D \ell_p}\sim \frac{\alpha \delta_z}{\eta D}\frac{\Delta C}{l_p}\frac{L}{\sqrt N}\;,
\end{equation}
where $\tau_{\text{diff}} \sim \ell_N^2/D$ and $\tau_{\text{adv}} \sim \ell_N/U$. We further used the expression for $U$ and $\ell_N = \sqrt{L^2/N}$. As mentioned earlier, at higher $N$ surface tension gradient $\Delta C/l_p$ decreases ($\sim N^{-\beta}$ with $\beta>0$) resulting in weaker flow $U$ and reduced Péclet number.\\

This suggests that, for fixed $L, \eta, D$ and swimmer size $\ell_{p}$ the Péclet number (Pe) increases linearly with concentration ($C$) while decreases with swimmer number as $1/N^{\beta + 1/2}$.
We note that our derivation of the Péclet number is analogous to that of Boniface et al. \cite{boniface2021role}, the key difference being the inclusion of a confinement-induced length scale in the present case.\\

\begin{figure}[h!]
    \centering
    \includegraphics[width=1.0\linewidth]{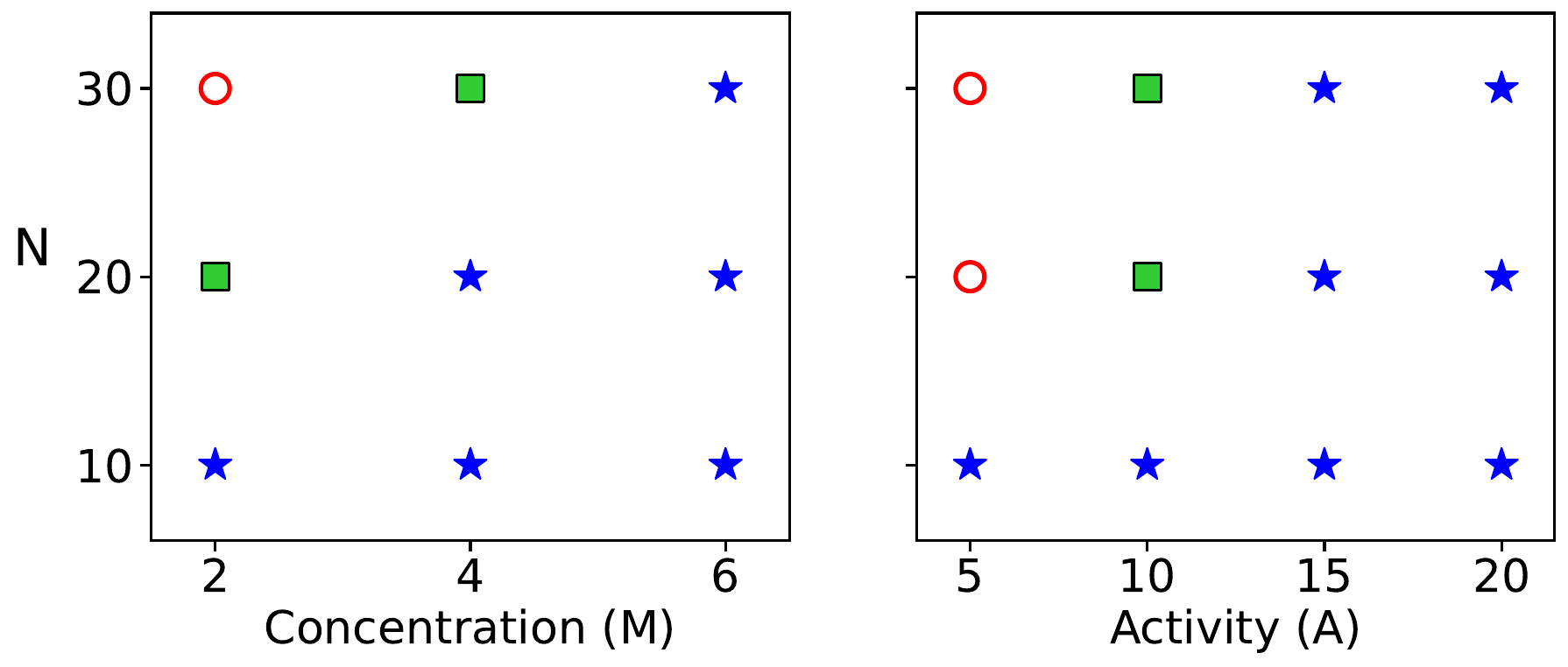}
    \caption{Phase-state diagrams obtained from experiments (left) and simulations (right). Circles, squares and stars represent disordered state ($\Phi<0.6$), moderately ordered state ($0.6<\Phi<0.9$) and fully ordered state ($\Phi\approx 0.99$), respectively. }
    \label{phase_state}
\end{figure}

For a fixed number of swimmers (e.g., $N=30$), increasing the concentration leads to an increase in the Péclet number. As a result, the system transitions from a disordered state to an ordered state. Conversely, if the concentration is kept constant (e.g., $2\,\mathrm{M}$) while increasing the number of swimmers $N$, the Péclet number decreases, leading to a transition from an ordered state to a disordered state. This behavior is consistent with observations in many active matter systems \cite{fins2024steer,caprini2024dynamical, chan2024interplay, moncho2012peclet}, where higher Péclet numbers favor the emergence of collective order, whereas lower Péclet numbers promote disordered dynamics. 


\textit{Model}. To understand the collective dynamics observed in the experiments, we adopt and extend a reduced model previously developed for camphor-driven systems\cite{kitahata2013spontaneous,koyano2026intrinsic,nagayama2023reaction,nagayama2004theoretical,kohira2001synchronized,hayashima2001camphor}, in which reaction–diffusion dynamics of the camphor field generate propulsion through surface-tension gradient. Interactions between swimmers, including alignment, excluded-volume effects \cite{liang2015research,smith1998geometric}, and boundary-induced reorientation, are incorporated phenomenologically. 
\begin{figure*}[htb!]
    \centering    \includegraphics[width=1.0\textwidth]{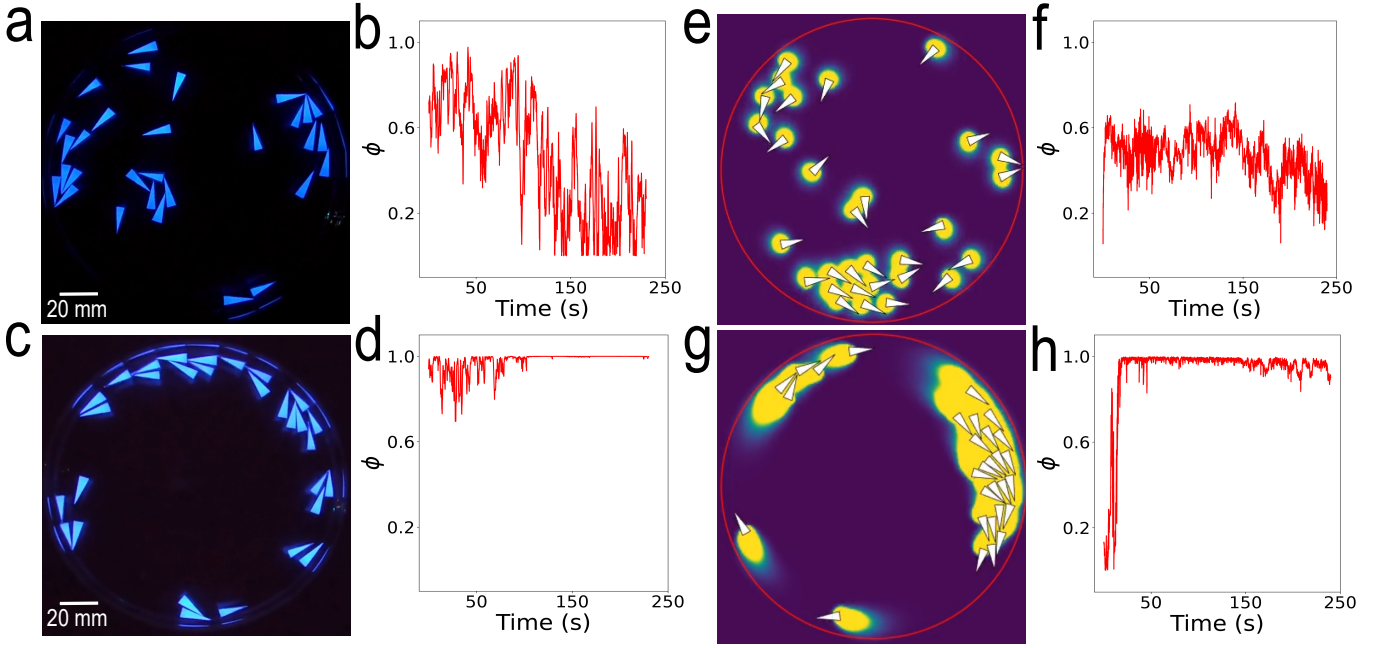}
    \caption{Snapshots from the experimental video and simulation video at the 110th second of 30 swimmers at concentration (a) 2M, (c) 6M and activity parameter (e) A=5  and (g) A=15. (b), (d), (f) and (h) shows the time evolution of the order parameter $\Phi$ corresponding to (a), (c), (e) and (g), respectively.  
    }
    \label{fig:fig2}
\end{figure*}
We write coupled equations for the camphor concentration 
field $c({\bf r},t)$ at the water surface and the motion of the swimmers ${\bf r}_i(t)$. Here
$c({\bf r},t)$ follows reaction diffusion dynamics (Eq.\ref{eq:camphor_pde} below), with a source term $S({\bf r},t)$ which integrates over the release of camphor from the painted regions 
$\mathcal C_i$ of the $i$-th swimmer.   
\begin{equation}
\label{eq:camphor_pde}
\frac{\partial c(\mathbf r,t)}{\partial t}
=D \nabla^2 c(\mathbf r,t)-k\, c(\mathbf r,t)
+S(\mathbf r,t).
\end{equation}
Here $D$ and $k$ are the diffusion coefficient
and the decay rates, respectively, and 
\begin{equation}
\label{eq:source}
S(\mathbf r,t)=
\sum_{i=1}^{N}
\sum_{r_{i\alpha} \in \mathcal C_i}
A_{\mathrm{de}}(t)\,
\delta\!\left(\mathbf r-\mathbf r_{i\alpha}\right),
\end{equation}
The activity  $A_{de}(t)=A(0)\mathrm e^{-t/\tau}$ is assumed to decay exponentially in time.
The translational and rotational motions of the swimmers, in viscous media, are assumed to be overdamped due to their light weight and relatively large size. The Langevin equations read,
\begin{equation}
 \dot{\mathbf{r}}_i
= \mu_t \mathbf{F}_i + {\eta}^t_i(t), \;  
\dot{\theta}_i
= \mu_r \left[\sum_{j\neq i} \tau^{\mathrm{align}}_{ij}
+ \tau^{\mathrm{wall}}_i
\right]
+ \eta^r_i(t),
\end{equation}
where $\mu_t$ and $\mu_r$ are translational and rotational mobilities, and  
${\eta}^t_i$ and $\eta^r_i$ are translational and rotational white Gaussian noise (see section VIII of SI for all parameter values). The force 
${\bf F}_i$ produces unidirectional motion along the instantaneous long axis of the swimmer, and is computed from the local surface tension
$\gamma(c,\mathbf r,t) =\gamma_0 -\frac{A_{\mathrm{de}}(t)}{N}\, c(\mathbf r,t)$,
which is reduced from its base value $\gamma_0$ due to the release of camphor. Instead of computing the surface tension gradient to get the Marangoni force ${\bf F}_i$ over the full body of the swimmer we follow a phenomenological formula
$
\mathbf{F}_i
=
\oint_{\mathcal{C}_i}
\gamma\!\left(c\right)
\,\hat{\mathbf{n}}_i\,\mathrm{d}\ell
$, employed in previous studies \cite{kitahata2013spontaneous,koyano2026intrinsic,nagayama2023reaction,nagayama2004theoretical,kohira2001synchronized,hayashima2001camphor,masanta2026dynamics}. 
Here $\hat{\mathbf n}_{i}$ is the outward normal to the coated region $\mathcal C_i$ (see SI section V (D)).
Instead of explicitly including shape dependent hydrodynamic interaction and Cheerios effect, which dictate swimmer-swimmer and swimmer-wall interactions, we use effective torques $\tau_{ij}^{align}$ and $\tau_i^{wall}$ (details in SI section V (E and I)). The excluded volume interaction between the swimmers was implemented using separating axes theorem (SAT) \cite{liang2015research, smith1998geometric}, see SI section V (H).

The full phase states obtained from both simulations (Movie6 and Movie7) and experiments are shown in Fig.\ref{phase_state}, demonstrating the agreement between the experimental observations and the model. This minimal model reproduces the essential experimental observations: boundary-following motion, transient alignment, and the activity and density-dependent transitions between disordered, ordered flocking states (Fig.\ref{fig:fig2} and
SI-Fig.8).


\par
\noindent


 In most theoretical models binary alignment is imposed as a rule (as also in our minimal model) and the development of long range dynamic correlation, leading to flocking, is the emergent feature. 
Our experiments, instead, highlights appropriate swimmer shape and hydrodynamic interaction to be a necessary criteria for binary alignment and its stability. Interestingly, in bird flocks and fish shoals an individual bird or fish maintains a pointed front, similar to our triangular swimmers, in the direction of movement. 



In addition, the $N$ dependent order to disorder re-entrant transition, that we found, is also novel and our dimensional analysis
argument suggests that confinement brings 
a new length scale $l_N$ into consideration. This re-entrant transition is qualitatively similar to the transition to glassiness at high density, found in Ref\cite{pagonobaraga2024}. However, unlike our case, where hydrodynamics is responsible, the origin of the re-entrant transition was different in Ref\cite {pagonobaraga2024}.

\textit{Acknowlegements.} We sincerely thank Amitabha Nandi and Pabitra Masanta for their useful suggestions and discussions on modeling this system. A.T. and S.K. acknowledge the financial assistance from the IITB.  T.K. acknowledges financial support from IITB in the form of a Post-Doctoral Fellowship. N. K. acknowledges financial support from DST-SERB for CRG Grant No. CRG/2020/
002925 and IITB for the seed grant. P.P. acknowledges financial support from the SERB, Department of Science and Technology (DST) (No.CRG/2022/001541), India. A.S. acknowledges the Science and Engineering Research Board (SERB), India (Project Nos. CRG/2019/005944) for financial support.

\bibliographystyle{apsrev4-1}
\bibliography{ref}  



\end{document}